# GENERAL FRAMEWORK FOR THE PROBABILISTIC DESCRIPTION OF EXPERIMENTS

E.R. LOUBENETS

*MaPhySto[1], Department of Mathematical Sciences, University of Aarhus,
Ny Munkegade, DK-8000, Aarhus C, Denmark*
*E-mail: elena@imf.au.dk*

**Abstract**
We introduce a new mathematical framework for the probabilistic description of an experiment on a system of any type in terms of information representing this system initially. Based on the notions of an information state and a generalized observable, this framework allows us to subsume different types of randomness and experiment effects within a single mathematical structure. Adjusting this framework to the quantum case, we clarify what is really "quantum" in quantum measurement theory.

## 1 Introduction

The problem of the relation between the statistical model of quantum theory and the formalism of probability theory is a point of intensive discussions, beginning from von Neumann's axioms [1] in quantum measurement theory and Kolmogorov's axioms [2] in the theory of probability.

In the physical literature on quantum physics one can find statements on the peculiarities of "quantum" probabilities and "quantum" events. In the mathematical physics literature, the structure of probability theory associated with the formalism of random variables is often referred to as *classical* probability or Kolmogorov's model.

From the mathematical point of view, *classical* probability is embedded as a particular case into "non-commutative probability theory" - the algebraic framework (see [3] and references therein), based on the structure of the statistical model of quantum theory. However, since the algebraic framework does not cover the description of all possible general probabilistic situations, this framework cannot be considered as an extension of probability theory.

---

[1] A Network in Mathematical Physics and Stochastics, funded by The Danish National Research Foundation.

Moreover, by its structure, the algebraic framework stands also aside from the developments in quantum measurement theory connected with the operational approach [4-8], the contextual approach [9] and the quantum stochastic approach [10-12]. The latter developments are more close to the spirit of probability theory.

The aim of this presentation is:
- to introduce the basics of a new mathematical framework [13] for the probabilistic description of an experiment on a system of any type in terms of information space representing this system initially;
- to answer the question "what is really quantum in quantum measurement theory" and what follows from the specification of a new probabilistic formalism to the quantum case.

The new framework incorporates the main probabilistic concepts of *classical* probability, *quantum* probability, quantum measurement theory and some general statistical models and allows us to subsume different types of randomness and experiment effects within a single mathematical structure.

The model of *classical* probability and the statistical model of quantum theory are both included into this new framework as particular cases.

In the quantum case, the new framework clarifies the origin and the structure of the basic notions introduced in quantum measurement theory axiomatically.

## 2 Probabilistic framework

### 2.1 General settings

Consider the general scheme of an experiment $E$ upon a system $S$ of any type:

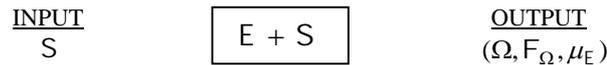

INPUT: $S$   E + S   OUTPUT: $(\Omega, \mathsf{F}_\Omega, \mu_E)$

Here:
$\Omega$ is a set of observed outcomes $\omega$;
$\mathsf{F}_\Omega$ is an algebra of subsets of $\Omega$, representing mathematically possible questions on an outcome $\omega$ being posed under this experiment;
$\mu_E$ is the probability distribution of outcomes under this experiment.

*According to Kolmogorov's axioms* [2], $\mathsf{F}_\Omega$ is a $\sigma$-algebra and $\mu_E$ is a $\sigma$-additive probability measure on a measurable space $(\Omega, \mathsf{F}_\Omega)$.

In probability theory, a measurable space $(\Omega, \mathsf{F}_\Omega)$ specifying the results of an experiment is called an outcome space while a measure space $(\Omega, \mathsf{F}_\Omega, \mu_E)$ is referred to as an outcome probability space.

Consider the following initial value problem:
- *To express $\mu_E$ in terms of properties, characterizing a system* S *before an experiment*.

The correct solution of this problem must automatically point out situations where properties, characterizing a system initially, provide "*no knowledge*" on the probabilistic description of an experiment. This "*no knowledge*" case means the independence of the probability distribution $\mu_E$ on the variations of properties, specifying a system initially.
To solve this problem, let us express mathematically in a very general setting the initial specification of a system

*2.2 Information state*

Denote by:
$\theta$ - a collection of properties of any nature, characterizing a system initially;
$\Theta$ - the set of all possible collections $\theta$. No linear structure of $\Theta$ is, in general, assumed;
$\mathsf{F}_\Theta$ - a $\sigma$-algebra of subsets of $\Theta$, specifying our knowledge on collections $\theta$. Specifically, we consider the case where the uncertainty of possible $\theta$ is represented on $(\Theta, \mathsf{F}_\Theta)$ by a $\sigma$-additive probability measure $\pi$: $\mathsf{F}_\Theta \to [0,1]$.

The above mathematical setting is rather general and covers a broad class of probabilistic situations under an experiment, in particular:
- in *classical* probability;
- in *quantum* probability;
- in all those situations where each $\theta$ is interpreted as representing a "bit" of information on a system and a measure $\pi$ describes statistical weights of possible bits.

Due to its informational context, any finite positive real-valued measure $\tilde{\pi}$ on $(\Theta, \mathsf{F}_\Theta)$, differing from a probability distribution $\pi$ only by normalization, represents the same initial information on a system $S$. We denote this equivalence class of measures by $[\pi]$ and introduce the following notion [13].

**Definition 1** *We call a triple* $(\Theta, \mathsf{F}_\Theta, [\pi])$ *an **information state**.*

We refer to a measurable space $(\Theta, \mathsf{F}_\Theta)$ as a system ***information space*** and note that, in general, $(\Theta, \mathsf{F}_\Theta)$ can be of any nature.

## 2.3 Generalized observable

In view of the introduced notion of an information state, the considered initial value problem can be represented by the following scheme:

| INPUT | | OUTPUT |
|---|---|---|
| $S$ in a state $(\Theta, \mathsf{F}_\Theta, [\pi])$ | E + S | $(\Omega, \mathsf{F}_\Omega, \mu_\mathsf{E}(\cdot\,;[\pi]))$ |

Due to the informational context of a probability distribution $\pi$, we assume the validity of the following statistical axiom (see also in [14,15]).

**Statistical axiom**
*For any experiment* $\mathsf{E}$*, the mapping* $\mu_\mathsf{E}(\cdot\,;\cdot)$ *satisfies the relation*:
(1) $\quad \mu_\mathsf{E}(B\,;[\pi]) = \alpha_1 \mu_\mathsf{E}(B\,;[\pi_1]) + \alpha_2 \mu_\mathsf{E}(B\,;[\pi_2]), \quad \forall B \in \mathsf{F}_\Omega,$
*for any* $\pi = \alpha_1 \pi_1 + \alpha_2 \pi_2 \quad \alpha_1, \alpha_2 > 0, \quad \alpha_1 + \alpha_1 = 1.$

In these general settings, we prove [13] the following theorem[2].

**Theorem 1**
*For a mapping* $\mu_\mathsf{E}(\cdot\,;\cdot)$*, satisfying (1), the function*
$$\Phi(\cdot\,;\cdot): \quad \mathsf{F}_\Omega \times \Theta \to [0,1],$$
*defined by the relation*[3]
$$\Phi(B;\theta) := \mu_\mathsf{E}(B;[\delta_\theta]), \quad \forall B \in \mathsf{F}_\Omega, \quad \forall \theta \in \Theta,$$
*has the properties*[4]:

---

[2] The case where a set $\Theta$ is finite and a $\sigma$-algebra $\mathsf{F}_\Theta$ is of the special type was considered in [6].
[3] Here, $\delta_\theta, \forall \theta \in \Theta,$ is a Dirac measure on $(\Theta, \mathsf{F}_\Theta)$.

(i) $\Phi(\cdot;\theta):\ \mathsf{F}_\Omega \to [0,1]$ *is a $\sigma$-additive probability measure for any $\theta \in \Theta$*;

(ii) $\Phi(B;\cdot):\ \Theta \to [0,1]$ *is a measurable function for any $B \in \mathsf{F}_\Omega$*;

*and, for any probability distribution $\pi$ on $(\Theta,\mathsf{F}_\Theta)$,*

(2) $$\mu_\mathsf{E}(B;[\pi]) = \int_\Theta \Phi(B;\theta)\,\pi(d\theta), \qquad \forall B \in \mathsf{F}_\Omega.$$

Notice that the representation (2) is not an assumption of a statistical model. In the general setting of theorem 1, this representation is proved without any extra probabilistic restrictions.

We can now interpret the derived representation (2) in two different ways:

(a) To an experiment $\mathsf{E}$ upon a system $\mathsf{S}$, there corresponds the unique *random element* $f$ on $(\Theta,\mathsf{F}_\Theta)$, defined by

$$(f(\theta))(B) := \Phi(B;\theta), \qquad \forall B \in \mathsf{F}_\Omega;\ \forall \theta \in \Theta,$$

that is, with values $f(\theta),\ \forall \theta \in \Theta$, that are $\sigma$-additive scalar probability measures on $(\Omega,\mathsf{F}_\Omega)$;

(b) To an experiment $\mathsf{E}$ on $\mathsf{S}$,, there corresponds the normalized $\sigma$-additive measure

$$\Pi:\ \mathsf{F}_\Omega \to B_M^{(+)}(\Theta),\qquad \Pi(\Omega) = \mathsf{I}_{B(\Theta)},$$

defined uniquely by the relation:

$$(\Pi(B))(\theta) := \Phi(B;\theta),\qquad \forall B \in \mathsf{F}_\Omega;\ \forall \theta \in \Theta.$$

Here,

$$\mathsf{I}_{B(\Theta)}(\theta) = 1,\quad \forall \theta \in \Theta,$$

and we denote by $B(\Theta)$ the Banach space of all bounded complex valued functions on $\Theta$ and by $B_M^{(+)}(\Theta)$ the set of all measurable positive bounded real valued functions on $\Theta$.

Thus, $\Pi$ is a normalized $\sigma$-additive measure with values $\Pi(B),\ \forall B \in \mathsf{F}_\Omega$, which are measurable bounded positive real-valued functions on $(\Theta,\mathsf{F}_\Theta)$.

---

[4] In probability theory, a function on $\mathsf{F}_\Omega \times \Theta$, with the properties (i), (ii), is called a Markov kernel.

We take the second variant since the properties of a positive mapping-valued measure $\Pi$ are similar to those of a scalar probability measure. This choice allows us:
- to formalize the description of any experiments, in particular, on classical and quantum systems, within a single mathematical framework;
- to formalize, in the most general settings, the description of an effect of a non-destructive experiment on a system;
- to formalize, in the most general settings, the description of joint and consecutive experiments;
- to specify the case of a classical measurement. Due to its set-up, the latter experiment does not perturb a system.

Notice that the notion of a normalized positive mapping-valued measure $\Pi$ is similar to that of a normalized positive operator-valued (POV) measure in quantum measurement theory, where the latter is often referred to as a generalized observable.

In the most general settings, we introduce the following notion [13].

**Definition 2** *We call a normalized $\sigma$ - additive measure*
$$\Pi: \quad \mathsf{F}_\Omega \to \mathrm{B}_M^{(+)}(\Theta), \qquad \Pi(\Omega) = \mathsf{I}_{B(\Theta)},$$
*a **generalized observable**, with outcome space $(\Omega, \mathsf{F}_\Omega)$ and on the information space $(\Theta, \mathsf{F}_\Theta)$.*

Each value $\Pi(B)$, $B \in \mathsf{F}_\Omega$, of a generalized observable $\Pi$ is a measurable bounded positive real-valued function on an information space $(\Theta, \mathsf{F}_\Theta)$.

The set of all generalized observables, with given outcome space, is convex linear.

*2.4 Representation theorem*

The notion of an *information state* incorporates the generality of a probability space in probability theory.

The introduced notion of a *generalized observable* covers the description of:
- a random variable – in *classical* probability;
- an observable - in *quantum* probability;
- a POV measure - in quantum measurement theory.

In the introduced terminology, theorem 1 is formulated in the following way [13].

**Representation Theorem**

*To an experiment* $\mathsf{E}$ *with outcomes in* $(\Omega, \mathsf{F}_\Omega)$, *performed on a system* $\mathsf{S}$, *represented initially by an information space* $(\Theta, \mathsf{F}_\Theta)$, *there corresponds a unique generalized observable* $\Pi$ *on* $(\Theta, \mathsf{F}_\Theta)$, *with the outcome space* $(\Omega, \mathsf{F}_\Omega)$, *such that*

(3) $$\mu_\mathsf{E}(B; [\pi]) = \int_\Theta (\Pi(B))(\theta)\, \pi(d\theta), \quad \forall B \in \mathsf{F}_\Omega, \quad \forall \pi.$$

***Remark 1.*** The converse statement is not true. A generalized observable on $(\Theta, \mathsf{F}_\Theta)$ does not necessarily represent some experiment on $\mathsf{S}$ and the same generalized observable $\Pi$ may, in general, correspond a variety of experiments on $\mathsf{S}$, which have the same probability distribution of outcomes.

Experiments on $\mathsf{S}$ represented by the same generalized observable $\Pi$ on $(\Theta, \mathsf{F}_\Theta)$ are statistically equivalent and we further denote by $[\Pi]$ the equivalence class of these experiments.

***Remark 2.*** Let information spaces $(\Theta, \mathsf{F}_\Theta)$ and $(\Theta', \mathsf{F}_{\Theta'})$ be isomorphic, that is, there exists a bijection $f: \Theta' \to \Theta$ with functions $f$ and $f^{-1}$ being measurable. In this case, there is the one-to-one correspondence:
$$\pi(F) = \pi'(f^{-1}(F)), \quad \forall F \in \mathsf{F}_\Theta,$$
between the information states $(\Theta, \mathsf{F}_\Theta, [\pi])$ and $(\Theta', \mathsf{F}_{\Theta'}, [\pi'])$.

From (3) it follows:
$$\mu_\mathsf{E}(B; [\pi]) = \int_{\Theta'} (\Pi(B))(\theta)\, \pi(d\theta)$$
$$= \int_{\Theta'} (\Pi(B) \circ f)(\theta')\, \pi'(d\theta') = \mu'_\mathsf{E}(B; [\pi']), \quad \forall B \in \mathsf{F}_\Omega.$$

Hence, on isomorphic information spaces $(\Theta', \mathsf{F}_{\Theta'})$ and $(\Theta, \mathsf{F}_\Theta)$, the same experiment $\mathsf{E}$ is described by generalized observables $\Pi'$ and
$$\Pi(B) = \Pi'(B) \circ f^{-1}, \quad \forall B \in \mathsf{F}_\Omega,$$
respectively.

*2.5 Image generalized observable*

A generalized observable $\Pi$ on $(\Theta, \mathsf{F}_\Theta)$, describing an experiment on $\mathsf{S}$, cannot be, in general, referred to a system property, existing before an experiment.

However, this is the case if $\Pi$ has a special structure:

(4) $\qquad (\Pi_{im}(B))(\theta) = \chi_{f^{-1}(B)}(\theta), \qquad \forall \theta \in \Theta, \quad \forall B \in \mathsf{F}_\Omega,$

where:

(i) $\chi_F(\theta)$ is an indicator function of a subset $\forall F \in \mathsf{F}_\Theta$;

(ii) $f : \Theta \to \Omega$ is a measurable function (random element - in the terminology of probability theory);

(iii) $f^{-1}(B) = \{\theta \in \Theta : f(\theta) \in B\}$ is the preimage in $\mathsf{F}_\Theta$ of a subset $B \in \mathsf{F}_\Omega$.

We call a generalized observable of the type (4) an *image generalized observable*. For an experiment, described on $(\Theta, \mathsf{F}_\Theta)$ by an image generalized observable, the probability distribution of outcomes is given by

$$\mu_E(B; [\pi]) = \pi(f^{-1}(B)), \qquad \forall B \in \mathsf{F}_\Omega, \quad \forall \pi,$$

and, hence, an image generalized observable corresponds to a classical "errorless" measurement (see [15, 13]) on defining the values of a property $f$ on a system initial information space $(\Theta, \mathsf{F}_\Theta)$. This property existed before an experiment.

## *2.6 On initial uncertainty*

In our settings, the randomness may be, in general, caused by:
- the uncertainty, encoded in an information state where, in general, two mathematical objects – the elements of a set $\Theta$ and a measure $\pi$ are responsible for that;
- a probabilistic set-up of an experiment itself.

To clarify the first point, consider the situation where a system initial information state $(\Theta, \mathsf{F}_\Theta, [\pi])$ is *induced* (see [13]) by an information state $(\Theta', \mathsf{F}_{\Theta'}, [\pi'])$. This means that there exists a unique generalized observable $\mathsf{S}$ on $(\Theta', \mathsf{F}_{\Theta'})$, with the outcome space $(\Theta, \mathsf{F}_\Theta)$, such that

(5) $\qquad \pi(F) = \int_{\Theta'} (\mathsf{S}(F))(\theta') \, \pi'(d\theta'), \qquad \forall F \in \mathsf{F}_\Theta.$

Consider an experiment $\mathsf{E}$, with outcomes in $(\Omega, \mathsf{F}_\Omega)$, performed on a system $\mathsf{S}$ in a state $(\Theta, \mathsf{F}_\Theta, [\pi])$. Let $\Pi$ be a generalized observable representing this experiment on $(\Theta, \mathsf{F}_\Theta)$. Then, due to (3) and (5), the outcome probability distribution is given by

(6) $$\mu_E(B;[\pi]) = \int_{\Theta'} (\Pi'(B))(\theta')\, \pi'(d\theta'),$$

where
$$(\Pi'(B))(\theta') := \int_{\Theta} (\Pi(B))(\theta)\, (S(d\theta))(\theta')$$

is the generalized observable representing an experiment $E$ on the information space $(\Theta', F_{\Theta'})$.

Suppose now that $S$ is an image generalized observable:
$$(S(F))(\theta') = \chi_{\phi^{-1}(F)}(\theta'), \quad \forall \theta' \in \Theta', \quad \forall F \in F_{\Theta},$$

with $\phi: \Theta' \to \Theta$ being a measurable function.

Then, for any $F \in F_{\Theta}$, $B \in F_{\Omega}$,
$$\pi(F) = \pi'(\phi^{-1}(F)), \qquad \Pi'(B) = \Pi(B) \circ \phi.$$

and, in (6),
$$\mu_E(B;[\pi]) = \mu'_E(B;[\pi'])$$
$$= \int_{\Theta'} (\Pi(B) \circ \phi)(\theta')\, \pi'(d\theta'),$$

Even if initial $\pi' = \delta_{\theta'}$ is a Dirac measure, the probability distribution
$$\mu'_E(B;[\delta_{\theta'}]) = (\Pi(B) \circ \phi)(\theta')$$

is, in general, not deterministic and depends on the type of the further experiment $E$. Only if $\Pi$ is also an image generalized observable:
$$(\Pi(B))(\theta) = \chi_{\psi^{-1}(B)}(\theta), \quad \forall \theta \in \Theta,$$

then the probability distribution of outcomes
$$\mu_E(B;[\pi]) = \pi'(f^{-1}(B)), \quad f = \psi \circ \phi, \quad \forall B \in F_{\Omega},$$

is an image of the probability measure $\pi'$ in an information state $(\Theta', F_{\Theta'}, [\pi'])$ and the result of the experiment $E$ is certain whenever initial $\pi'$ is a Dirac measure.

Thus, the uncertainty in elements $\theta \in \Theta$, which is induced by a probability distribution on some underlying information space, can be "lifted" in a "non-perturbing" manner only under experiments of the image type.

*2.7 "No knowledge" case*

In general, an initial information space of a system may not provide any knowledge for the probabilistic prediction on an experiment.

For example, if we take an initial information space of a quantum system in terms of a finite-dimensional complex Hilbert space $C^N$ then this initial information space does not provide any knowledge on experiments for defining a momentum (or position) of this quantum system.

Similarly, if an initial information space of a quantum system is based on an infinite-dimensional Hilbert space, which does not have any internal structure, then this initial information space cannot provide any knowledge for experiments, connected with internal degrees of freedom of this quantum system.

**Proposition 1**

*An information space* $(\Theta, \mathsf{F}_\Theta)$ *of* S *provides* ***"no knowledge"*** *on an experiment upon* S *iff this experiment is represented on* $(\Theta, \mathsf{F}_\Theta)$ *by a generalized observable*

(7) $\quad\quad\quad\quad \Pi(B) = \nu(B) I_{B(\Theta)}, \quad \forall B \in \mathsf{F}_\Omega,$

*where* $\nu(\cdot): \mathsf{F}_\Omega \to [0,1]$ *is a probability distribution on* $(\Omega, \mathsf{F}_\Omega)$.

We call a generalized observable ***trivial*** *if it has* the form (7).

From proposition 1 it follows that a trivial information space (where $\mathsf{F}_\Theta = \{\varnothing, \Theta\}$) provides "no knowledge" on experiments upon a system S. A non-trivial information state space provides knowledge only on the class of experiments, which are represented on $(\Theta, \mathsf{F}_\Theta)$ by non-trivial generalized observables.

### 2.8  *Joint generalized observables*

Let $\Pi$ be a generalized observable on $(\Theta, \mathsf{F}_\Theta)$ describing a joint experiment with outcomes in a product space $(\Omega_1 \times \Omega_2, \mathsf{F}_{\Omega_1} \otimes \mathsf{F}_{\Omega_2})$.

The marginal generalized observables of $\Pi$, defined by the relations:

(8) $\quad \Pi_1(B_1) := \Pi(B_1 \times \Omega_2), \quad\quad \Pi_2(B_2) := \Pi(\Omega_1 \times B_2)$,

describe experimental situations where the outcomes in $(\Omega_2, \mathsf{F}_{\Omega_2})$ and $(\Omega_1, \mathsf{F}_{\Omega_1})$, respectively, are ignored completely.

With respect to $\Pi_1$ and $\Pi_2$, the generalized observable $\Pi$, satisfying (8), is called *joint.*

To any generalized observables $\Pi_1$ and $\Pi_2$ on $(\Theta, \mathsf{F}_\Theta)$, there exists the unique generalized observable $\Pi$ on $(\Theta, \mathsf{F}_\Theta)$, with the product outcome space $(\Omega_1 \times \Omega_2, \mathsf{F}_{\Omega_1} \otimes \mathsf{F}_{\Omega_2})$, such that
(9) $\quad\quad\quad\quad (\Pi(B_1 \times B_2))(\theta) = (\Pi(B_1))(\theta)(\Pi(B_2))(\theta),$
for any $B_1 \in \mathsf{F}_{\Omega_1}, B_2 \in \mathsf{F}_{\Omega_2}, \theta \in \Theta$.
We call a generalized observable $\Pi$ of the form (9) *product* and denote it by
(10) $\quad\quad\quad\quad\quad\quad \Pi = \Pi_1 \times \Pi_2 .$

Let, for a system of some concrete type, any experiment is described by a generalized observable on $(\Theta, \mathsf{F}_\Theta)$ which satisfies some super-selection rule. This rule specifies the definite class of generalized observables on $(\Theta, \mathsf{F}_\Theta)$.

For any two generalized observables $\Pi_1$, $\Pi_2$ in this class, the product generalized observable $\Pi_1 \times \Pi_2$ may not belong to this class and, hence, cannot represent an experiment on this concrete system.

## 3  Description of non-destructive experiments

Consider now the case where, immediately after an experiment, a system $\mathsf{S}$ can be specified in terms of an information state. We call this experiment *non-destructive*.
In general, "input" and "output" information spaces of $\mathsf{S}$ may be different.

### *3.1  Extended generalized observable*

Let, immediately before and after a non-destructive experiment $\mathsf{E}$ with outcomes in $(\Omega, \mathsf{F}_\Omega)$, a system $\mathsf{S}$ be represented by information spaces $(\Theta_{in}, \mathsf{F}_{\Theta_{in}})$ and $(\Theta_{out}, \mathsf{F}_{\Theta_{out}})$, respectively.
Consider the general scheme of a non-destructive experiment:

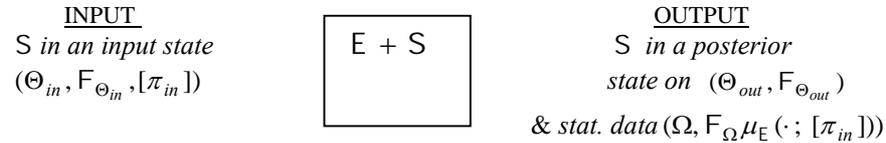

Each trial of a non-destructive experiment results in a compound outcome

$$(\omega, \theta_{out}) \in \Omega \times \Theta_{out},$$

and we assume that the probability distribution $\nu_\mathsf{E}(\cdot\,;[\pi_{in}])$ of these compound outcomes satisfies the statistical axiom (1).

Then, according to the representation theorem, this non-destructive experiment is represented on $(\Theta_{in}, \mathsf{F}_{\Theta_{in}})$ by the unique generalized observable $\mathrm{Y}$ with the product outcome space

$$(\Omega \times \Theta_{out}, \mathsf{F}_\Omega \otimes \mathsf{F}_{\Theta_{out}}).$$

For a generalized observable $\mathrm{Y}$, the marginal generalized observable

$$\mathrm{M}_Y(B) := \mathrm{Y}(B \times \Theta_{out}), \quad \forall B \in \mathsf{F}_\Omega,$$

defines the probability distribution of outcomes in $(\Omega, \mathsf{F}_\Omega)$:

(11) $$\mu_\mathsf{E}(B\,;[\pi_{in}]) = \int_{\Theta_{in}} (\mathrm{M}_Y(B))(\theta_{in})\, \pi_{in}(d\theta_{in}), \quad \forall B \in \mathsf{F}_\Omega,$$

while the marginal generalized observable

$$\mathrm{S}_Y(F_{out}) := \mathrm{Y}(\Omega \times F_{out}), \quad \forall F_{out} \in \mathsf{F}_{\Theta_{out}},$$

determines the unconditional probability distribution

(12) $$\tau_\mathrm{Y}(F_{out}\,;[\pi_{in}]) = \int_{\Theta_{in}} (\mathrm{S}_Y(F_{out}))(\theta_{in})\, \pi_{in}(d\theta_{in}), \quad \forall F_{out} \in \mathsf{F}_{\Theta_{out}},$$

of posterior system outcomes $\theta_{out} \in \Theta_{out}$.

We call $\mathrm{Y}$, $\mathrm{M}_Y$, $\mathrm{S}$ as *extended*, *outcome* and *system* generalized observables, respectively, on an information space $(\Theta_{in}, \mathsf{F}_{\Theta_{in}})$.

## *3.2 Experiment effects*

Let a non-destructive experiment $\mathsf{E}$ be represented on an initial information space $(\Theta_{in}, \mathsf{F}_{\Theta_{in}})$ of $\mathsf{S}$ by a non-trivial extended generalized observable $\mathrm{Y}$.

Under each experimental trial, consider an *effect* of this experiment on a system. We specify this effect in terms of a posterior information state [13], conditional on an outcome event $B \in \mathsf{F}_\Omega$.

The conditional measure

(13) $$\pi_Y^{out}(F_{out} \mid B; \pi_{in}) = \frac{\int_{\Theta_{in}} (\mathrm{Y}(B \times F_{out}))(\theta_{in})\, \pi_{in}(d\theta_{in})}{\int_{\Theta_{in}} (\mathrm{M}_Y(B))(\theta_{in})\, \pi_{in}(d\theta_{in})}$$

defines the probability that, immediately after a single trial where only the event that the outcome $\omega \in B$ has been recorded, the posterior system outcome $\theta_{out}$ belongs to a subset $F_{out} \in \mathsf{F}_{\Theta_{out}}$.

Hence, immediately after this single trial, the triple

(14) $\quad\quad\quad\quad\quad (\Theta_{out}, \mathsf{F}_{\Theta_{out}}, [\pi_Y^{out}(\cdot \mid \Omega; \pi_{in})])$

represents the conditional posterior information state.
In view of the above structure (13) of a conditional posterior information state, we introduce the following notion.

For any bounded real-valued measure $\nu$ on $(\Theta_{in}, \mathsf{F}_{\Theta_{in}})$ and any subsets $B \in \mathsf{F}_\Omega$, $F_{out} \in \mathsf{F}_{\Theta_{out}}$, let denote by

(15) $\quad\quad\quad\quad (\mathsf{M}_Y^{\inf}(B)\nu)(F_{out}) := \int_{\Theta_{in}} (Y(B \times F_{out}))(\theta_{in}) \nu(d\theta_{in})$

a real-valued measure on $(\Omega \times \Theta_{out}, \mathsf{F}_\Omega \otimes \mathsf{F}_{\Theta_{out}})$.

The mapping $\mathsf{M}_Y^{\inf}$, defined by (15), is a measure on $(\Omega, \mathsf{F}_\Omega)$ with values $\mathsf{M}_Y^{\inf}(B)$, $\forall B \in \mathsf{F}_\Omega$, which are positive bounded linear operators from the Banach space of $\sigma$-additive bounded real-valued measures on $(\Theta_{in}, \mathsf{F}_{\Theta_{in}})$ to the Banach space of $\sigma$-additive bounded real-valued measures on $(\Theta_{out}, \mathsf{F}_{\Theta_{out}})$.

The vector-valued measure $\mathsf{M}_Y^{\inf}$ is normalized - in the sense that $(\mathsf{M}_Y^{\inf}(\Omega)\nu)(\Theta_{out}) = 1$ whenever $\nu(\Theta_{in}) = 1$.

**Definition 3** *We call the mapping $\mathsf{M}_Y^{\inf}$, defined by (15) to an extended generalized observable $Y$, **an information state instrument**.*

From definition 3 and the formulae (11), (13) it follows that, under a non-destructive experiment, the probability distribution of outcomes and the conditional posterior states are given, respectively, by

(16) $\quad\quad \mu_{\mathsf{E}}(B; [\pi_{in}]) = (\mathsf{M}_Y^{\inf}(B)\pi_{in})(\Theta_{out}), \quad \forall B \in \mathsf{F}_\Omega, \quad \forall \pi_{in}$,

and

(17) $\quad\quad\quad \pi_Y^{out}(F_{out} \mid B; \pi_{in}) = \dfrac{(\mathsf{M}_Y^{\inf}(B)\pi_{in})(F_{out})}{\mu_{\mathsf{E}}(B; [\pi_{in}])}$,

$\forall B \in \mathsf{F}_\Omega, \forall F_{out} \in \mathsf{F}_{\Theta_{out}}, \forall \pi_{in}$.

Hence, for a non-destructive experiment, the information state instrument $\mathsf{M}_Y^{\text{inf}}$ defines both - the probability distribution of outcomes and the family of conditional posterior information states.

***Remark 1.*** The above formulae, arising in our framework in the most general settings automatically are introduced axiomatically in the frame of the operational approach (see [4-8] and the review sections in [10-12]).

***Remark 2.*** The conditional change

$$(\Theta_{in}, \mathsf{F}_{\Theta_{in}}, [\pi_{in}]) \mapsto (\Theta_{out}, \mathsf{F}_{\Theta_{out}}, [\pi_Y^{out}(\cdot \mid B; \pi_{in})]),$$

given by (17), represents, in the most general settings, the ***phenomena of "reduction" of an information state*** following a single experimental trial.

In the frame of our approach it is clearly seen that this phenomena is, in general, caused:
- by the "renormalization" of information on a system conditioned on the recorded event under a single experimental trial;
- by the "dynamical" change of an information state of a system in the course of an experiment.

### 3.3 *Description of consecutive experiments*

In a very general setting, let consider the description of a consecutive experiment. Suppose that initially a system $\mathsf{S}$ is an information state $(\Theta_{in}, \mathsf{F}_{\Theta_{in}}, [\pi_{in}])$ and the first experiment $\mathsf{E}_1$, with outcomes in $(\Omega_1, \mathsf{F}_{\Omega_1})$, is described on $(\Theta_{in}, \mathsf{F}_{\Theta_{in}})$ by an extended generalized observable $\mathsf{Y}_1$. Let

$$(\Theta_1, \mathsf{F}_{\Theta_1}, [\pi_{Y_1}^{out}(\cdot \mid B_1; [\pi_{in}])])$$

be a system posterior state immediately after a single experimental trial of an experiment $\mathsf{E}_1$ and a second experiment $\mathsf{E}_2$, with outcomes in $(\Omega_2, \mathsf{F}_{\Omega_2})$, be described on the information space $(\Theta_1, \mathsf{F}_{\Theta_1})$ by an extended generalized observable $\mathsf{Y}_2$.

For a consecutive experiment $\mathsf{E}_2 \circ \mathsf{E}_1$, the information state instrument has the form:

(18) $\qquad \widetilde{\mathsf{M}}^{\text{inf}}(B_1 \times B_2) = \mathsf{M}_2^{\text{inf}}(B_2) \circ \mathsf{M}_1^{\text{inf}}(B_1),$

for any subsets $B_1 \in \mathsf{F}_{\Omega_1}, B_2 \in \mathsf{F}_{\Omega_2}$.

Hence, from (15) and (16) it follows that the outcome generalized observable of a consecutive experiment is given by:

(19) $\quad (\widetilde{M}(B_1 \times B_2))(\theta_{in}) = \int_{\Theta_1} (M_{Y_2}(B_2))(\theta_1)(Y_1(B_1 \times d\theta_1))(\theta_{in})$,

for any $\theta_{in} \in \Theta_{in}$, $B_1 \in \mathsf{F}_{\Omega_1}$, $B_2 \in \mathsf{F}_{\Omega_2}$.

*3.4 Non-perturbing experiments*

Consider the description of a non-destructive experiment, described by a *product extended* generalized observable of the form:

(20) $\quad (Y(B \times F_{out}))(\theta_{in}) = (M_Y(B))(\theta_{in})(S_Y(F_{out}))(\theta_{in})$,

$\forall B \in \mathsf{F}_\Omega$, $\forall F_{out} \in \mathsf{F}_{\Theta_{out}}$, $\forall \theta_{in} \in \Theta_{in}$.

Suppose that, before an experiment,

$$\pi_{in}(F_{in}) = \delta_a(F_{in}) \quad \forall F_{in} \in \mathsf{F}_{\Theta_{in}},$$

with some $a \in \Theta_{in}$. Then, from (13) it follows that, for a product generalized observable (20),

$$\pi_Y^{out}(F_{out} \mid B; [\delta_a]) = (S_Y(F_{out}))(a),$$

for any $B \in \mathsf{F}_\Omega$, $F_{out} \in \mathsf{F}_{\Theta_{out}}$, $a \in \Theta_{in}$.

Let, for example, all atom subsets $\{\theta_{in}\}$ belong to a $\sigma$-algebra $\mathsf{F}_{\Theta_{in}}$ and, before an experiment, a system be in a pure information state. Then, under any experimental trial, the behaviour of this system does not depend on an outcome event, that is, on whether or not this system is observed.

We call a generalized observable of the form (20) **non-perturbing**.

A general form of a non-perturbing generalized observable Y is considered in [13].

Notice that, under a non-perturbing experiment, a posterior information state of a system may be, in general, different from an initial one and this change depends on the form of a system generalized observable $S_Y$.

## 4 Statistical description

We call any mapping on the set $\{(\Theta, \mathsf{F}_\Theta, [\pi]): \forall \pi\}$ of all information states a *statistical information state.*

In general, the notion of a statistical information state is less informative than the notion of an information state.

Consider the case where a set $\Theta$ represents a bounded subset of some Banach space $V$:

(21) $$\Theta = \{\theta \in V: \quad \|\theta\|_V \leq C\}$$

and a $\sigma$-algebra $\mathsf{F}_\Theta \supseteq \mathsf{B}_\Theta$, where $\mathsf{B}_\Theta$ is the trace on $\Theta$ of the Borel $\sigma$-algebra on the Banach space $V$.

In this case, we specify the notion of a *mean information state* [13].

**Definition 4** *We call* $\eta([\pi]) = \int_\Theta \theta \pi(d\theta) \in V$ *a* ***mean information state*** *on* $(\Theta, \mathsf{F}_\Theta)$.

The set $\mathsf{R}$ of all mean information states is convex linear.

*4.1 Conditional posterior mean states*

Suppose that under a non-destructive experiment both system information sets $\Theta_{in}$ and $\Theta_{out}$ satisfy the condition (21), possibly with $V_{in} \neq V_{out}$.

Let Y be an extended generalized observable, describing this experiment on $(\Theta_{in}, \mathsf{F}_{\Theta_{in}})$, and $\eta_{in}([\pi_{in}])$ be an initial mean information state of a system.

The formula
$$\eta_{out}([\pi_{in}]|B) = \int_{\Theta_{out}} \theta_{out} \pi_Y^{out}(d\theta_{out}|B; \pi_{in}), \quad \forall B \in \mathsf{F}_\Omega,$$

represents the conditional average over posterior system outcomes $\theta_{out} \in V_{out}$ and we refer to it as a ***conditional posterior mean information state*** [13].

Due to (17), we have the following expression for a conditional posterior mean information state

(22) $$\eta_{out}([\pi_{in}]|B) = \frac{\int_{\Theta_{out}} \theta_{out} (\mathsf{M}_Y^{\inf}(B)\pi_{in})(d\theta_{out})}{\mu_Y(B;[\pi_{in}])}.$$

Under the *complete statistical description* of a non-destructive experiment, we mean the knowledge, for each initial information state, of the outcome probability distribution and the family
$$\{\eta_{out}([\pi_{in}]|B); \ \forall B \in \mathsf{F}_\Omega\}$$
of all conditional posterior mean information states.

In general, for different initial information states, which induce the same initial mean information state $\eta_{in}$, the conditional posterior mean information states, corresponding to the same event $B \in \mathsf{F}_\Omega$, may be different.

However, for the class of generalized observables, which we proceed to introduce, the mapping

$$\eta_{in} \mapsto \eta_{out}(\eta_{in} \mid B)$$

is well defined.

### 4.2 Pre-linear extended generalized observables

Suppose that on each of the Banach spaces $V_{in}$ and $V_{out}$ there exists a continuous linear functional[5] $\mathsf{l}_\gamma$ such that

$$\mathsf{l}_\gamma[\theta_\gamma] = 1, \quad \forall \theta_\gamma \in \Theta_\gamma.$$

Here, for short, we use index $\gamma$ to denote "*in*" or "*out*". We also denote by $\mathsf{R}_\gamma$ the set of all mean information states $\eta_\gamma$ on $(\Theta_\gamma, \mathsf{F}_{\Theta_\gamma})$.

Let a generalized observable Y' on $(\Theta_{in}, \mathsf{F}_{\Theta_{in}})$ be such that, in (22), the mapping

$$(23) \qquad (\mathsf{M}_Y^{st}(B))(\theta_{in}) = \int_{\Theta_{out}} \theta_{out}(Y'(B \times d\theta_{out}))(\theta_{in}), \quad \forall B \in \mathsf{F}_\Omega,$$

has a *unique convex linear* extension (in $\theta_{in}$) to all of the set $\mathsf{R}_{in} \supseteq \Theta_{in}$ of initial mean information states.

In this case, for any subset $B \in \mathsf{F}_\Omega$, both, the outcome probability distribution

$$(24) \qquad \begin{aligned} \mu_\mathsf{E}(B;[\pi_{in}]) &= \mathsf{l}_{out}[\int_{\Theta_{in}} \theta_{out}(Y'(B \times d\theta_{out}))(\theta_{in}) \pi_{in}(d\theta_{in})] \\ &= \mathsf{l}_{out}[(\mathsf{M}_{Y'}^{st}(B))(\eta_{in})] = \tilde{\mu}_\mathsf{E}(B;\eta_{in}) \end{aligned}$$

and the conditional posterior state

---

[5] In the quantum case considered in section 5, this functional is given by $tr[\cdot]$.

(25)
$$\eta_{out}([\pi_{in}] \mid B) = \frac{\int_{\Theta_{out}} \theta_{out}(Y(B \times d\theta_{out}))(\eta_{in})}{\tilde{\mu}_E(B; \eta_{in})}$$

$$= \frac{(M^{st}_{Y_{prl}}(B))(\eta_{in})}{\tilde{\mu}_E(B; \eta_{in})} = \tilde{\eta}_{out}(\eta_{in} \mid B)$$

*depend only on an initial mean information state* $\eta_{in}$, but not on an initial information state.

We refer to an extended generalized observable with the property (23) ***pre-linear*** and, for concreteness, denote it by $Y_{prl}$.

For a pre-linear generalized observable $Y_{prl}$, we call the corresponding mapping $M^{st}_{Y_{prl}}$, *induced* by the convex linear extension of (23), a ***mean information state instrument***[6].

Due to (24) and (25), $M^{st}_{prl}$ gives the complete statistical description of the corresponding non-destructive experiment and, moreover, specifies the conditional change of a mean information state:
$$\eta_{in} \mapsto \eta_{out}(\eta_{in} \mid B).$$

*Under an experiment* $E$, *described by a pre-linear extended generalized observable, the probability distribution of outcomes depends only on an initial mean information state and satisfies the relation*

(26) $\quad \tilde{\mu}_E(B; \eta_{in}) = \alpha_1 \tilde{\mu}_E(B; \eta^{(1)}_{in}) + \alpha_2 \tilde{\mu}_E(B; \eta^{(2)}_{in}), \quad \forall B \in F_\Omega,$

*for any* $\eta_{in} = \alpha_1 \eta^{(1)}_{in} + \alpha_2 \eta^{(2)}_{in}, \ \alpha_1, \alpha_2 > 0, \ \alpha_1 + \alpha_1 = 1.$

***Example.*** Denote by $V^*$ the Banach space of continuous linear functionals on $V$. Let $(\Lambda, F_\Lambda)$ be a measurable space. Consider a normalized measure
$$\Pi : F_\Lambda \to V^*, \quad \Pi(\Lambda) = I,$$

---

[6] The mean information state instrument $M^{st}_Y$ is a measure with values $M^{st}_Y(B), \quad \forall B \in F_\Omega$, which are bounded linear operators $R_{in} \to V_{out}$.

with values that are *continuous linear* functionals on $V$. Any such measure represents a generalized observable on $(\Theta, \mathsf{F}_\Theta)$, specified by (21).
We call this generalized observable (extended, outcome or system) as ***linear***.
Any extended linear generalized observable is pre-linear.

Notice that the notion of a mean information state instrument can be introduced only in the case where information sets have a Banach space structure and, for experiments, represented by pre-linear (in particular, linear) extended generalized observables.

## 5 Generalized Quantum Measurements

Let now specify the notions, introduced in sections 2-4, to the case of experiments on a quantum system (see [13], for details).
Further, by a generalized quantum measurement we mean an experiment on a quantum system resulting in imprints of any nature in the classical world.

### 5.1 Quantum information state

Let a quantum system $\mathsf{S}_q$ be described in terms of a separable complex Hilbert space $\mathsf{H}$. Denote by $\mathsf{T}_\mathsf{H}$ the Banach space of all trace class operators on $\mathsf{H}$ and by $\mathsf{L}_\mathsf{H}$ and $\mathsf{L}_\mathsf{H}^{(+)} \subset \mathsf{L}_\mathsf{H}$ the Banach space of bounded linear operators on $\mathsf{H}$ and the set of positive bounded linear operators on $\mathsf{H}$.
For a quantum system, we take *an information space* to be represented by [7]
$$(\mathsf{P}_\mathsf{H}, \mathsf{B}_{\mathsf{P}_\mathsf{H}})$$
where:
$\mathsf{P}_\mathsf{H}$ is the set of all one-dimensional projections $p = |\psi\rangle\langle\psi|$ on $\mathsf{H}$;
$\mathsf{B}_{\mathsf{P}_\mathsf{H}}$ is the trace on $\mathsf{P}_\mathsf{H}$ of the Borel $\sigma$-algebra on $\mathsf{T}_\mathsf{H}$.

A mean information state on $(\mathsf{P}_\mathsf{H}, \mathsf{B}_{\mathsf{P}_\mathsf{H}})$ is given by
$$\rho = \int_{\mathsf{P}_\mathsf{H}} p\, \pi(dp)$$

---

[7] We could also take a quantum information space to be represented by the set $\mathsf{R}_\mathsf{H}$ of all density operators on $\mathsf{H}$.

and represents a density operator on $H$. The set of all mean information states of a quantum system coincides with the set $R_H$ of density operators on $H$.

According to the terminology accepted in quantum theory, we further refer to a quantum mean information state $\rho$ on $(P_H, B_{P_H})$ as a quantum state, pure or mixed.

*5.2 Quantum outcome generalized observable*

Since $P_H$ is a bounded subset of the Banach space $T_H$ and the positive bounded linear functional $tr[\cdot]: T_H \to C$ represents a functional $l_\gamma$, specified in section 4.2, all items of the *statistical description*, which we discuss in a very general setting in section 4, are valid in the quantum case.

*In quantum measurement theory, the relation (26) is assumed to be valid for any quantum measurement.*

**Theorem 2**
*For any quantum measurement, the outcome generalized observable $M^{(q)}$ is linear and is given by*:
$$(M^{(q)}(B))(T) = tr[TM(B)], \quad \forall T \in T_H, \ \forall B \in F_\Omega,$$
*where*
$$M: F_\Omega \to L_H^{(+)}$$
*is a normalized positive operator-valued (POV) measure on $(\Omega, F_\Omega)$.*

*The probability distribution of outcomes on $(\Omega, F_\Omega)$ depends only on an initial quantum mean information state (i.e. a quantum state $\rho$) and has the form*:
$$\mu(B; \rho) = tr[\rho M(B)].$$

*5.3 Quantum extended generalized observable*

Consider now a non-destructive quantum measurement.
Let immediately after this measurement a quantum system be described in terms of a Hilbert space $K$.

In quantum measurement theory, a conditional posterior quantum state is *postulated to have the form*[8]

(27) $$\rho_{out}(\rho_{in} \mid B) = \frac{(\mathsf{M}_q(B))(\rho_{in})}{\mu(B; \rho_{in})},$$

with the mapping $\mathsf{M}_q$, called a quantum state instrument and representing an operation-valued measure on $(\Omega, \mathsf{F}_\Omega)$, satisfying the relation

$$tr[(\mathsf{M}_q(\Omega))(T)] = 1, \quad \forall T \in \mathsf{T}_\mathsf{H}.$$

*In the light of our framework* (see section 4.2), the relation (27) implies:

- any non-destructive quantum measurement is described by a pre-linear extended generalized $\mathrm{Y}_{prl}^{(q)}$ on $(\mathsf{P}_\mathsf{H}, \mathsf{B}_{\mathsf{P}_\mathsf{H}})$;

- a quantum state instrument $\mathsf{M}_q$ represents the mean information state instrument $\mathsf{M}^{st}$ (see (23)).

Moreover, any non-destructive quantum measurement is of the perturbing type (see section 3.3).
Consider a linear extended generalized observable $\mathrm{Y}_{lin}^{(q)}$ on $(\mathsf{P}_\mathsf{H}, \mathsf{B}_{\mathsf{P}_\mathsf{H}})$.

**Theorem 3**
*A quantum linear extended generalized observable has the form:*
$$(\mathrm{Y}_{lin}^{(q)}(B \times F_{out}))(T) = tr[TY(B \times F_{out})],$$

*for any* $T \in \mathsf{T}_\mathsf{H}$, $B \in \mathsf{F}_\Omega$, $F_{out} \in \mathsf{B}_{\mathsf{P}_\mathsf{K}}$, *where* $Y$ *is a POV measure on*
$$(\Omega \times \mathsf{P}_\mathsf{K}, \mathsf{F}_\Omega \otimes \mathsf{B}_{\mathsf{P}_\mathsf{K}})$$
*with values that are positive bounded linear operators on* $\mathsf{H}$.
*The quantum state instrument, corresponding to a linear extended generalized observable, is given by*
$$(\mathsf{M}_q(B))(T) = \int_{\mathsf{P}_\mathsf{K}} p_{out}(\mathrm{Y}_{lin}^{(q)}(B \times dp_{out}))(T), \quad \forall T \in \mathsf{T}_\mathsf{H}, \forall B \in \mathsf{F}_\Omega,$$
*and is completely positive.*

---

[8] See [4-8] and the review sections in [10-12].

Recall that, in quantum measurement theory, complete positivity of a quantum state instrument is introduced axiomatically.

**Acknowledgements**

I am indebted to Ole E. Barndorff-Nielsen and Goran Peskir for valuable remarks and useful discussions.
The support, given by MaPhySto for the research reported here, is gratefully acknowledged.